\documentclass{PoS}

\newcommand{\be}{\begin{equation}}
\newcommand{\ee}{\end{equation}}
\newcommand{\bea}{\begin{eqnarray}}
\newcommand{\eea}{\end{eqnarray}}

\newcommand{\Pg}{\hat{{\rm P}}_{\rm G}}
\newcommand{\Tr}{\,\hbox{\rm Tr}}

\title{Symmetries and exponential error reduction in YM theories 
on the lattice: theoretical aspects and simulation results
\thanks{Combined contribution of the two parallel talks given by the authors}}

\ShortTitle{Symmetries and exponential error reduction in YM theories on the 
lattice}

\author{Michele Della Morte \\
       Institut f\"ur Kernphysik, University of Mainz,\\
       Johann-Joachim-Becher Weg 45, D-55099 Mainz, Germany\\
       E-mail: \email{morte@kph.uni-mainz.de}}

\author{Leonardo Giusti\\
       CERN, Physics Department, 1211 Geneva 23, Switzerland and \\
       Dipartimento di Fisica, Universit\'a di Milano Bicocca, Piazza \\
       della Scienza 3, I-20126 Milano, Italy\\
       E-mail: \email{leonardo.giusti@cern.ch}}

\abstract{
\vspace{-11.5cm}
The path integral of a quantum system with an exact symmetry can 
be written as a sum of functional integrals each giving the 
contribution from
quantum states with definite symmetry properties.
We propose a strategy to compute each of them,
normalized to the one with vacuum quantum numbers, by a Monte Carlo
procedure whose cost increases power-like with the time extent of
the lattice. This is achieved thanks to a
multi-level integration scheme, inspired by the transfer matrix
formalism, which exploits the symmetry and the locality in time
of the underlying statistical system. As a result the cost of computing
the lowest energy level in a given channel, its multiplicity
and its matrix elements is exponentially reduced with respect to the
standard path-integral Monte Carlo. We briefly illustrate the approach in 
the simple case of the one-dimensional harmonic oscillator and discuss 
in some detail its extension to the four-dimensional Yang Mills theories.
We report on our recent new results in the SU(3) Yang--Mills theory on 
the relative contribution to the partition function of the parity-odd states.
\vspace{-17.5cm}
\begin{flushright}
MKPH-T-09-20\\
CERN-PH-TH/2009-189
\end{flushright}
}

\FullConference{The XXVII International Symposium on Lattice Field Theory - LAT2009\\
		 July 26-31 2009\\
		 Peking University, Beijing, China}

\begin{document}

\section{Introduction}
\subsection{The problem}
\noindent Lattice field theories can be studied numerically by Monte Carlo simulations. 
They allow to address non-perturbative problems from first principles, and 
for most of the theories the lattice provides the only known non-perturbative 
definition. The mass of the lowest states in a given channel can, for instance, 
be extracted from the Euclidean time-dependence of a suitable two-point correlation 
function. Very often, however, the statistical error of the Monte Carlo
estimate grows exponentially with time, and in practice
it is not possible to find a window where statistical and systematic
errors are both under control. The problem is easily explained by looking
at the states contributing asymptotically in time to the two-point 
function and to its variance. Whenever the energy of the asymptotic state
in the variance is smaller than twice that in the two-point function, 
the noise to signal ratio is going to grow exponentially in 
time~\cite{Parisi:1983ae,Lepage:1989hd}. 
The standard Monte Carlo approach fails basically because for any given  
field configuration all asymptotic states of the theory are allowed 
to propagate in the time direction, regardless of the quantum numbers of 
the source fields. Their contributions disappear in the Monte Carlo average
for the two-point function but sum up in the noise. As shown in the
following subsection, the issue is already there for a simple system such 
as the harmonic oscillator.  We use the latter to introduce the basic ideas 
of the method that was proposed in \cite{DellaMorte:2007zz,DellaMorte:2008jd}, 
a ``symmetry-constrained'' Monte Carlo, and to show how it 
avoids the exponential increase of the signal-to-noise ratio.
\subsection{The case of the harmonic oscillator}
\noindent We consider the one-dimensional harmonic oscillator on the lattice. 
We recall here a few basic equations. More details can be found 
in~\cite{DellaMorte:2007zz} to which we refer for any 
unexplained notation. The system is described by the Hamiltonian
\begin{equation}
\hat H = \frac{\hat p^2}{2m} + V(\hat x) \quad {\rm with} \quad 
V(\hat x)=\frac{1}{2} m \omega^2 \hat x^2\; .
\end{equation}
This operator is invariant under parity transformations, therefore its 
eigenstates can be classified according to a parity quantum number ($+$ or
$-$). We label the corresponding energy levels as ${\cal E}_i^+$ 
and ${\cal E}_j^-$ respectively. 
The transfer operator between two consecutive time slices is 
defined as
\begin{equation}
\displaystyle
\hat {\cal T} = e^{-\frac{a}{2} V(\hat x)}\,
e^{-a \frac{\hat p^2}{2m}}\, e^{-\frac{a}{2} V(\hat x)}\; ,
\end{equation}
an its matrix elements in the coordinate basis 
\begin{equation}
\displaystyle
\langle x_{n+1}| \hat {\cal T} | x_{n}\rangle
\equiv \left(\frac{m}{2\pi a}\right)^{1/2} T_{n+1,n}
\end{equation}
can be computed explicitly
\begin{equation}
T_{n+1,n} = e^{-a L_{n+1,n}}\; ,
\end{equation}
with
\begin{equation}
L_{n+1,n} \equiv {\cal L}(x_{n+1},x_n) = 
\frac{m}{2}\left(\frac{x_{n+1}-x_n}{a}\right)^2 +
\frac{V(x_{n+1})}{2} + \frac{V(x_n)}{2} \; .
\end{equation}

The statistical variance
associated to the two-point correlation function 
$\langle x_l x_k \rangle$ (interpolating parity odd states) is
\begin{equation}
\sigma^2 = \langle x^2_l x^2_k \rangle  - \langle x_l x_k \rangle^2 \; ,
\end{equation}
and, at asymptotically large time separations, the signal-to-noise
ratio can be easily computed in the underlying quantum field theory
\begin{equation}\displaystyle
\frac{\langle x_l x_k \rangle}
     {\sigma} =
\frac{|\langle {\cal E}_0^-|\hat x | {\cal E}_0^+\rangle|^2}
     {|\langle {\cal E}_0^+|\hat x^2 | {\cal E}_0^+\rangle|}\,
     e^{-a\, ({\cal E}_0^- - {\cal E}_0^+)|l-k|} + \cdots
\end{equation}
The ratio decreases exponentially in time, as announced. As we will describe 
in the following the problem here can be solved by introducing the ``sign'' 
and the ``module'' fields.

We define the complete set of parity eigenstates
\begin{equation}
|x,\pm\rangle = \frac{1}{\sqrt{2}}(|x\rangle \pm |-x\rangle)\; ,\qquad
 \hat {\cal P}|x,\pm\rangle = \pm |x,\pm\rangle\; .
\end{equation}
The invariance of the Hamiltonian under parity implies
\begin{equation}
\displaystyle
 \langle s',x_{n+1}| \hat {\cal T} | x_{n},s\rangle = 
\left(\frac{2 m}{\pi a}\right)^{1/2}
T^s_{n+1,n}\, \delta_{s's}\; ,
\end{equation}
with
\begin{eqnarray}
T^s_{n+1,n} & = & \frac{1}{2}\,
e^{-a \Lambda^+_{n+1,n}} \left\{
e^{a \Lambda^-_{n+1,n}} +s\,  e^{-a \Lambda^-_{n+1,n}}\right\}\; ,\\[0.25cm]
\Lambda^\pm_{n+1,n} & = & \frac{1}{2}\Big\{{\cal L}(-x_{n+1},x_n)
\pm  {\cal L}(x_{n+1},x_n)\Big\}\; ,
\end{eqnarray}
and the functional integral can be written as
\begin{equation}
Z = \sum_{s=\pm} Z^s\; , \qquad Z^s = \int \prod_{n=0}^{N-1} d x_n\, 
T^s_{n+1,n}\; ,
\end{equation}
where $N$ is the extent of the lattice.
We further define
\begin{equation}
T^+_{n+1,n} \equiv e^{-a L^+_{n+1,n}}  =
e^{-a \Lambda^+_{n+1,n}}
\cosh\{\,a \Lambda^-_{n+1,n}\}\; ,
\end{equation}
and cast the functional integrals in the form
\begin{equation}
Z^+ = \int \prod_{n=0}^{N-1} d x_n\, e^{-S^+}\;, \qquad
Z^- = \int \prod_{n=0}^{N-1} d x_n \, e^{-S^+} \prod_{m=0}^{N-1}
\tanh\{a\, \Lambda^-_{m+1,m}\}\; ,
\end{equation}
where $S^+ \equiv a\sum_{n=0}^{N-1} L^+_{n+1,n}$.
The path integral is thus rewritten as a sum of two functional
integrals giving the contribution from parity even and odd states
respectively. Each integrand is a product of transfer matrix elements
between quantum states with definite parity.
The two-point correlation function reads ($k<l$)
\begin{eqnarray}
\langle x_l x_k \rangle & = &\frac{1}{Z} \int \prod_{n=0}^{N-1} d x_n 
\, e^{-S^+}
\Big\{
x_l \prod_{m=k}^{l-1} \tanh\{a\, \Lambda^-_{m+1,m}\}x_k \nonumber\\
& + &
\prod_{m=l}^{N-1} \tanh\{a\, \Lambda^-_{m+1,m}\} x_l x_k
\prod_{m=0}^{k-1} \tanh\{a\, \Lambda^-_{m+1,m}\} \Big\}\; .
\end{eqnarray}
Each term is now  the expectation value of a factorized observable in a system
described by the action $S^+$. The expression reminds of the factorized form
used for the correlator of Polyakov loops in the pure gauge theory
in~\cite{Luscher:2001up}.
As done there a multi-level integration scheme can be introduced for
the system and the observable discussed here.
The key ingredients are sub-lattice averages, i.e. averages computed 
by numerically integrating over the degrees of freedom in a thick time-slice
of the lattice with the variables at the boundaries kept fixed, and the
recursive relations, which allow to obtain averages on large thick time-slices
as the product of those on smaller ones integrated over their boundaries
configurations. Both properties are due to the locality of the 
action. We do not repeat here the details concerning the construction
of the algorithm, they can be found in~\cite{DellaMorte:2007zz}
but rather report on the main results.

\begin{figure}
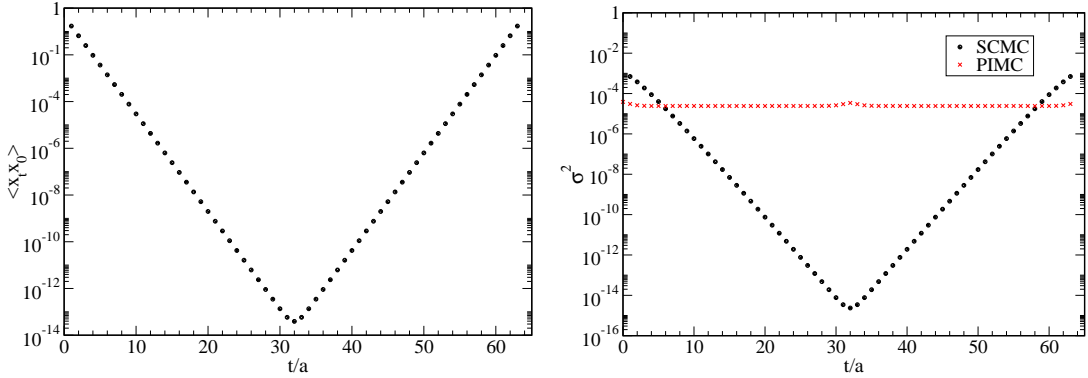

\includegraphics[width=7.0cm]{hs6_corr_64.eps}
\hspace{0.25cm}\includegraphics[width=7.0cm]{hs6_err_64.eps}
\caption{Left: two-point correlation function versus the time distance 
$t/a$ of the sources.
Right: errors on the correlation function as obtained with the multi-level
algorithm (SCMC) and with the standard path integral Monte Carlo (PIMC).}
\label{fig:corrs}
\end{figure}

\begin{figure}
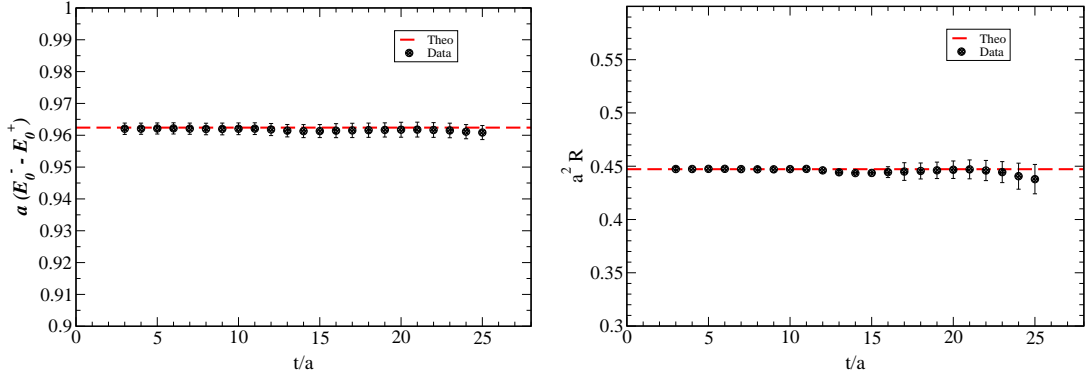

\includegraphics[width=7.0cm]{effmass_corr_64.eps}
\hspace{0.25cm}\includegraphics[width=7.0cm]{decay_corr_64.eps}
\caption{Left: effective energy difference extracted from
the two-point correlator at any time separation $t/a$.
Right: the ratio $R$ as defined in the text.}
\label{fig:effm_cor}
\end{figure}

In the left plot of Figure~\ref{fig:corrs} we show the two-point correlation
function computed on a lattice with $N=64$ points, with statistical errors 
being smaller than symbols. 
The error (SCMC) is shown on the right plot of the same Figure. 
The signal-to-noise ratio is depleted, as expected, (only) inversely 
proportional to the time distance of the sources. For comparison in the same 
plot it is also shown the statistical error obtained
with a standard Monte Carlo procedure (PIMC) which needed roughly the same CPU time.
It is clear that with our strategy the statistical error
is exponentially reduced, and at large time distances
it is lowered by many orders of magnitude. 
The effective energy-split $a\tilde\omega(t)$ extracted from the correlator 
is shown in the left plot of Figure~\ref{fig:effm_cor}, and it is in perfect
agreement with the theoretical expectation~\cite{Creutz:1980gp}. 
On the right plot of the same figure it is shown the effective estimate $R(t)$ 
of the square of the matrix  
element $\langle {\cal{E}}_0^- | \hat{x}| {\cal{E}}_0^+ \rangle$ computed as
\begin{equation}\displaystyle
\label{eq:Rn}
R(t) =\frac{\langle x_l x_k\rangle\; 
e^{\tilde\omega\frac{T}{2}}}{2 \cosh{\left[\tilde\omega
\left(\frac{T}{2}-a |l-k|\right)\right]}} \; ,
\end{equation}
which also agrees very well with the analytical result.

The approach described here is of inspiration for systems with a larger number
of degrees of freedom. In that case, however, one cannot make sense of the
``sign'' and ``module'' fields. New concepts have to be introduced as we 
illustrate in the following for the $SU(3)$ Yang-Mills theory \cite{DellaMorte:2008jd}.
\section{Extension to Yang-Mills theories on the lattice}
\noindent We again divide the parity even sector of the theory from the parity odd,
naming the energy levels $E_i^+$ and $E_j^-$ respectively.
The statistical variance of the estimate of a two-point
correlation function $\langle  O(x_0) O(0) \rangle$
of a parity-odd interpolating
operator $O$, computed by the standard Monte Carlo procedure, is defined as
\begin{equation}
\sigma^2 = \langle O^2(x_0)  O^2(0) \rangle  - \langle O(x_0) O(0) 
\rangle^2 \; .
\end{equation}
At asymptotically-large time separations the signal-to-noise
ratio  takes the form
\begin{equation}
\displaystyle
\frac{\langle  O(x_0) O(0) \rangle}
     {\sigma} =
\frac{|\langle E_1^-|\hat  O | 0 \rangle|^2}
     {|\langle 0 |\hat O^2 | 0 \rangle|}\,
     e^{- E_1^- x_0} + \cdots
\end{equation}
i. e. the signal is again depleted exponentially in time.

For the one-dimensional 
harmonic oscillator the quantity $\tanh\{a\, \Lambda^-_{m+1,m}\}$
represented the ratio $Z^-/Z^+$ for a system of one time-slice 
with fixed boundary configurations. Due to the regularity of the spectrum
this ratio is of O$(1)$. The same cannot be expected for systems with many
degrees of freedom, as the four-dimensional Yang-Mills theory, the ratio
will rather be of O$(e^{-(L/a)^3})$, with $L$ the spatial extent of the 
lattice.
However, if one considers  systems $d$ time-slices large, 
with $d\sim 1/T_{\rm c}$ and $T_{\rm c}$ the critical temperature, the
same ratio is now expected to be of O($e^{-E_1^-d}$) for each boundary 
configuration. These are the quantities we want to directly access and use
to rewrite our observables. To this end we first need to briefly recall 
the formalism of the transfer matrix, we refer to~\cite{DellaMorte:2008jd}
for a more thorough discussion. 
\subsection{Transfer matrix}
\noindent We adopt Wilson's regularization of gauge theories~\cite{Wilson:1974sk}.
The corresponding transfer matrix has been explicitly constructed 
in~\cite{Wilson:1977nj,Luscher:1976ms,Creutz:1976ch,Osterwalder:1977pc}. 
The functional integral with periodic boundary conditions in time 
can be written as
\begin{equation}\label{eq:Z1}
Z = \int \prod_{x_0=0}^{T-1}\,
{\rm \bf D}_3[V_{x_0}]\, {\rm T}\Big[V_{x_0+1},V_{x_0}\Big]
\end{equation}
where the transfer matrix elements among states $|V_{x_0}\rangle$ 
in the coordinate basis are defined as
\begin{equation}
{\rm T}\Big[V_{x_0+1},V_{x_0}\Big] = \int {\rm \bf D}[\Omega]
\; e^{-L[V^{\Omega}_{x_0+1},V_{x_0}]} \; ,
\end{equation}
with
\begin{equation}
L\Big[V_{x_0+1},V_{x_0}\Big] =  K\Big[V_{x_0+1},V_{x_0}\Big]
+  \frac{1}{2}W\Big[V_{x_0+1}\Big] + \frac{1}{2}W\Big[V_{x_0}\Big]\;,
\end{equation}
and $|V_{x_0}^\Omega \rangle$ is the result of a gauge transformation
$\Omega$ on the state $|V_{x_0}\rangle$.
The kinetic and the potential contributions to the Lagrangian are
given by
\be
K\Big[V_{x_0+1},V_{x_0}\Big] = \beta \sum_{{\bf x},k}
\left[ 1 - \frac{1}{3}{\rm Re}\Tr\left\{
V_{k}(x_0+1,{\bf x}) V^\dagger_{k}(x_0,{\bf x})
\right\}\right]\; ,
\ee
and
\be
W\Big[V_{x_0}\Big] = \frac{\beta}{2} \sum_{{\bf x}} \sum_{k,l}
\left[1 - \frac{1}{3}{\rm Re}\Tr\Big\{V_{kl}(x_0,{\bf x})\Big\}\right]\; ,
\ee
respectively, where $V_{kl}$ is the spatial plaquette computed with 
the links $V_k({\bf x})$. 
By exploiting the invariance of the Haar integration measure under
left and right multiplication it is easy to show that the transfer matrix
is gauge invariant. For a thick time-slice,
i.e. the ensemble of points in the sub-lattice with time
coordinates in a given interval $[x_0,y_0]$ and bounded by the equal-time
hyper-planes at times $x_0$ and $y_0$, the transfer matrix elements can be
introduced by the formula
\begin{equation}\label{eq:T1}
{\rm T}\Big[V_{y_0},V_{x_0}\Big] = \int \prod_{w_0=x_0+1}^{y_0-1}\,
{\rm \bf D}_3[V_{w_0}]\,
\prod_{z_0=x_0}^{y_0-1} {\rm T}\Big[V_{z_0+1},V_{z_0}\Big]\; .
\end{equation}

The parity transformation acts on the states in the coordinate basis as
\begin{equation}\label{eq:PT}
\mbox{ $\hat\wp$}\,|{\rm V}\rangle = |{\rm V}^\wp\rangle\; , \qquad
|{\rm V}\rangle = \Pg |V\rangle\; ,  \qquad V^\wp_k({\bf x}) = 
V^\dagger_k(-{\bf x}-\hat k)\; ,
\end{equation}
where $\Pg$ is the projector on gauge invariant states.
Again, we can then define a complete set of parity eigenstates 
\begin{equation}
|{\rm V},\pm\rangle = \frac{1}{\sqrt{2}}\Big[|{\rm V}\rangle \pm 
|{\rm V}^\wp\rangle \Big]\;, \qquad
\mbox{ $\hat\wp$}\, |{\rm V},\pm\rangle = \pm |{\rm V},\pm\rangle\; ,
\end{equation}
and their transfer matrix elements are given by
\begin{eqnarray}
\langle s', {\rm V}_{x_0+1}| \hat {\rm T} | {\rm V}_{x_0}, s\rangle & = & 2\,
\delta_{s's}\; {\rm T}^s\Big[V_{x_0+1},V_{x_0}\Big]\; ,\\[0.25cm]
{\rm T}^s\Big[V_{x_0+1},V_{x_0}\Big] & = & \frac{1}{2}
\left\{ {\rm T}\Big[V_{x_0+1},V_{x_0}\Big] + s\,
        {\rm T}\Big[V_{x_0+1},V^\wp_{x_0}\Big] \right\}\; .\label{eq:Ts}
\end{eqnarray}
For a thick time-slice the matrix elements between parity states
can be introduced by exploiting the same composition rule as in 
Eq.~(\ref{eq:T1}) with $\rm T$ replaced by ${\rm T}^s$.
In addition, the relations 
\bea
\int {\rm \bf D}_3[V_{z_0}]\,{\rm T}^s\Big[V_{y_0},V_{z_0}\Big]\;
{\rm T}^{-s}\Big[V_{z_0},V_{x_0}\Big] & = & 0\; ,\label{eq:comp2}\\[0.25cm]
\int {\rm \bf D}_3[V_{z_0}]\,{\rm T}^s\Big[V_{y_0},V_{z_0}\Big]\;
       {\rm T}\Big[V_{z_0},V_{x_0}\Big] & = &
{\rm T}^s\Big[V_{y_0},V_{x_0}\Big]\label{eq:comp3}
\eea
hold. 
In particular they imply that
\be\label{eq:Z3}
 \frac{{\rm T}^s[V_{y_0},V_{x_0}]}
      {{\rm T}\;[V_{y_0},V_{x_0}]}=
\frac{1}{Z_\mathrm{sub}}\int
 {\rm D}_4 [U]_\mathrm{sub} \; e^{-S[U]}\,
\frac{{\rm T}^s[U_{y_0},U_{y_0-1}]}
     {{\rm T}\,[U_{y_0},U_{y_0-1}]}\; ,
\ee
an useful expression for the practical implementation of the
multi-level algorithm described in the following.
The subscript ``sub'' indicates that the integral is performed over the
dynamical field variables in the thick time-slice $[x_0,y_0]$ with
the spatial components $U_k(x)$ of the boundary fields
fixed to $V_k(x_0,\vec {\bf x})$ and $V_k(y_0,\vec {\bf x})$ respectively.
Finally, by replacing ${\rm T}[V_{x_0+1},V_{x_0}]$  in Eq.~(\ref{eq:Z1})
by $\sum_s {\rm T}^s[V_{x_0+1},V_{x_0}]$
and repeatedly applying Eq.~(\ref{eq:comp2}), it is possible to rewrite the
path integral as a sum of functional integrals
\be
Z = \sum_{s=\pm} Z^s\; , \qquad Z^s = \int \prod_{x_0=0}^{T-1}\,
{\rm \bf D}_3[V_{x_0}]\, {\rm T}^s\Big[V_{x_0+1},V_{x_0}\Big]\; ,
\ee
each giving the contribution from gauge-invariant parity-even
and -odd states respectively.

The insertion of
${\rm T}^s [V_{y_0},V_{x_0}]$ in the path integral plays the r\^ole of
a projector, as on each configuration it allows the propagation in the
time direction of states with parity $s$ only. Indeed the parity transformation
of one of the boundary fields in ${\rm T}[V_{y_0},V_{x_0}]$
flips the sign of all contributions that it receives from the parity-odd
states while leaving invariant the rest. The very same applies
to the path integral in Eq.~(\ref{eq:Z1}) if the periodic boundary
conditions are replaced by $\wp$-periodic boundary conditions, i.e.
$V_T=V^\wp_0$. All contributions from the parity odd states are
then multiplied by a minus sign.
\subsection{The hierarchical integration scheme }
\noindent To determine the parity projector between two
boundary fields of a thick time-slice, the basic building block
to be computed is the ratio of transfer matrix elements
\be\label{eq:rat}
{\rm R}[V_{x_0+d},V_{x_0}] =
\frac{{\rm T}[V_{x_0+d},V^\wp_{x_0}]}{{\rm T}[V_{x_0+d},V_{x_0}]}\; .
\ee
As mentioned above, for $d$ of $O(1/T_{\rm c})$, 
the ratio $R$ is expected to be of $O(1)$. However
the integrands in the numerator and in the denominator on the r.h.s of
Eq.~(\ref{eq:rat}) are, in general, very different and the main
contributions to their integrals come from different regions of the phase 
space.
The most straightforward way for computing $R$ is to define a set of $n$
systems with partition functions ${\cal Z}_1\, \dots\, {\cal Z}_n$
designed in such a way that the relevant phase spaces of successive integrals
overlap and that ${\cal Z}_1={\rm T}[V_{x_0+d},V^\wp_{x_0}]$ and
${\cal Z}_n={\rm T}[V_{x_0+d},V_{x_0}]$. The ratio $R$ can then be calculated
as
\be
{\rm R} =
\frac{{\cal Z}_1}{{\cal Z}_2} \times \frac{{\cal Z}_2}{{\cal Z}_3}
\times \dots
\times \frac{{\cal Z}_{n-2}}{{\cal Z}_{n-1}}
\times \frac{{\cal Z}_{n-1}}{{\cal Z}_n} \; ,
\label{eq:interpol}
\ee
with each ratio on the r.h.s. being computable in a single
Monte Carlo simulation by averaging the proper reweighting factor.

\begin{figure}[htb]
\begin{center}
\includegraphics[angle=-90,width=8.0cm]{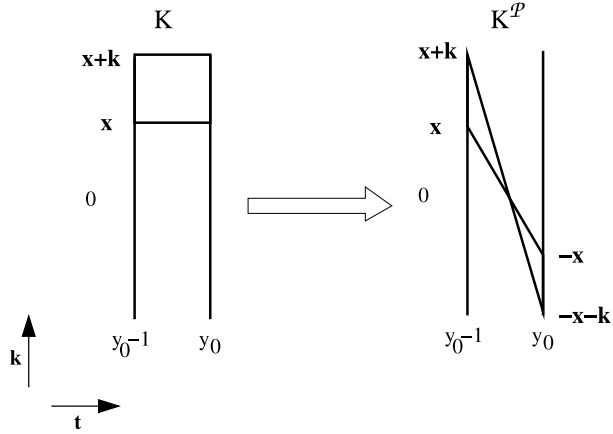}
\caption{Two dimensional representation of the plaquettes in the 
kinetic terms $K$ ($r=1/2$) and $K^\wp$ ($r=-1/2$)
on the time-slice $y_0-1$. The telescopic algorithm described in the text
bridges between the two systems in $L^3$ steps.}
\label{fig:KP}
\end{center}
\end{figure}

For the case at hand ${{\cal Z}_1}$ and ${{\cal Z}_n}$ are the partition
functions of two systems differing only for the boundary conditions in time.
In both cases Dirichlet boundary conditions are imposed but the boundary 
configurations at time $x_0 + d$ differ by a parity transformation. Instead of
relating the boundary configurations in such a way, we change the action of one
of the two systems on the last time-slice, by introducing a new temporal link
connecting the point ${\bold{x}},x_0+d-1$ on the last dynamical time-slice
to its parity transformed  $-{\bold{x}},x_0+d$ on the boundary.
We call the associated plaquette ``parity twisted'' space-time plaquette
and $K^\wp$ (parity twisted kinetic term) the sum of such plaquettes
(see Figure~\ref{fig:KP}).
To interpolate between the two systems we slowly switch off the 
coupling $\beta$
in $K$ by decreasing it in steps of size $\epsilon=1/L^3$ while increasing
the coupling in $K^\wp$ by the same amount, we distinguish the interpolating
actions by a parameter $-1/2 \leq r \leq 1/2$. In this way we move in $L^3$ steps 
from one system to the other. This means we need to perform a chain of $L^3$ 
Monte Carlo simulations within a Monte Carlo simulation and we therefore
have an algorithm, which scales with the second power of the volume $L^3$.
This is the case also for other known methods for computing ratios of partition
functions~\cite{Ferrenberg:1989ui,Hoelbling:2000su,deForcrand:2000fi}.

Once the projectors have been computed, the ratio of partition functions
$Z^s/Z$ can be calculated by implementing the hierarchical two-level
integration formula
\be\label{eq:bella}
\frac{Z^s}{Z} = \frac{1}{Z}\int {\rm D}_4 [U]\, e^{-S[U]}
\, {{\rm P}}^s_{m,d}\Big[T,0\Big]
\ee
where
${\rm P}^s_{m,d}\Big[y_0,x_0\Big]$ is defined as
\be
\displaystyle\label{eq:Psmd}
{{\rm P}}^s_{m,d}\Big[y_0,x_0\Big] =
{\prod_{i=0}^{m-1}}
\frac{{\rm T^s}[U_{x_0+(i+1)\cdot d},U_{x_0 + i\cdot d}]}
     {{\rm T}[U_{x_0+(i+1)\cdot d},U_{x_0 + i\cdot d}]}
\ee
with $m\ge 1$ and $y_0=x_0 + m \cdot d$. The procedure can, of course,
be generalized to a multi-level
algorithm. For a three-level one, for instance, each ratio on the
r.h.s of Eq.~(\ref{eq:Psmd}) can be computed by a
two-level scheme.
For each configuration of the boundary fields,
the magnitude of the product in our observable ${{\rm P}}^s_{m,d}[T,0]$
is proportional to $e^{-E^-_1\, T}$, and the statistical
fluctuations are reduced to this level.
This has to be  compared to the standard case in which 
each configuration gives a contribution to the signal which decreases
exponentially in time, whereas it contributes $O(1)$ to the noise (variance)
at any time distance. 
To achieve an analogous exponential gain
in the computation of the correlation functions, the projectors ${\rm T^s}$ 
have to be inserted in the proper way among the interpolating operators.
As a technical aside we remark that the computation of $\rm R$
requires a thermalization procedure for each value of $r$. We do not
expect the latter to be particularly problematic since, as mentioned above,
expectation values for consecutive values of $r$ refer to path integrals
with the relevant phase spaces which overlap. The ratio $\rm R$
is computed by simulating
systems corresponding to consecutive values of $r$ one after the other, and by
starting from the one used to extract the boundary fields ($r=1/2$).
\subsection{Results}
\noindent In the four-dimensional SU(3) Yang-Mills theory we have simulated lattices
with an inverse gauge coupling of $\beta=6/g^2_0=5.7$ which corresponds
to a value of the reference scale $r_0$ of about $2.93a$
\cite{Guagnelli:1998ud,Necco:2001xg}. The number of lattice points in each
spatial direction has been set to $L=6,8$ and $10$ corresponding to a linear
size of $1.0$, $1.4$ and $1.7$~fm respectively. For each spatial volume we have
considered several time extents $T$, the full list is
reported in Table~\ref{tab:lattices} together with the number of
configurations generated, the details of the multi-level
simulation algorithm used for each run and the results for $Z^-/Z$ and for 
the effective mass $M^-$ of the first parity-odd glueball state extracted from 
that ratio:
\be
M^-=-\frac{1}{T} \ln \left( \frac{Z^-}{Z}(T) \right) \;.
\label{eq:M-}
\ee
\begin{table}[htb]
\begin{center}
\setlength{\tabcolsep}{.85pc}
\begin{tabular}{|cccccccc|}
\hline
Lattice&$L$&$T$&$N_\mathrm{conf}$&$N_\mathrm{lev}$&$d$&${{Z^-}\over{Z}}
$&$aM^-$\\[0.125cm]
\hline
${\rm A}_1$&   6  &  4  & 50 & 2 & 4 & 0.409(8)& 0.223(5)\\[0.125cm]
${\rm A}_2$&      &  5  & 50 & 2 & 5 & 0.177(13)& 0.346(14)\\[0.125cm]
${\rm A}_3$&      &  6  & 50 & 2 & 6 & 0.069(7)& 0.446(17)\\[0.125cm]
${\rm A}_4$&      &  8  & 175& 2 & 4 & 1.47(28)$\cdot 10^{-2}$& 0.528(24)\\[0.125cm]
${\rm A}_5$&      & 10  & 50 & 2 & 5 & 2.2(5)$\cdot 10^{-3}$& 0.611(20)\\[0.125cm]
${\rm A}_6$&      & 12  & 90 & 2 & 6 & 6.6(17)$\cdot 10^{-4}$& 0.610(21)\\[0.125cm]
${\rm A}_7$&      & 16  & 48 & 2 & 8 & 2.8(8)$\cdot 10^{-5}$& 0.655(18)\\[0.125cm]
${\rm A}_8$&      & 20  & 48 & 3 & \{5,10\}&1.5(5)$\cdot 10^{-6}$&0.670(15)\\[0.125cm]
\hline
${\rm B}_1$&   8  &  4  & 20 & 2 & 4 &0.426(8)& 0.213(5)\\[0.125cm]
${\rm B}_2$&      &  5  & 25 & 2 & 5 &0.061(6)& 0.558(21)\\[0.125cm]
${\rm B}_3$&      &  6  & 75 & 2 & 3 &1.65(26)$\cdot 10^{-2}$& 0.685(27)\\[0.125cm]
${\rm B}_4$&      &  8  & 48 & 2 & 4 &1.37(26)$\cdot 10^{-3}$& 0.824(24)\\[0.125cm]
${\rm B}_5$&      &  12 & 48 & 3 & \{3,6\} &3.6(18)$\cdot 10^{-6}$& 1.045(41)\\[0.125cm]
${\rm B}_6$&      &  16 & 36 & 3 & \{4,8\} &5.2(19)$\cdot 10^{-8}$& 1.048(23)\\[0.125cm]
\hline
${\rm C}_1$&  10  &  4  & 20   & 2 & 4 &0.455(12)& 0.197(6)\\[0.125cm]
${\rm C}_2$&      &  5  & 24   & 2 & 5 &0.060(3)& 0.561(11)\\[0.125cm]
${\rm C}_3$&      &  6  & 50   & 2 & 3 &1.6(4)$\cdot 10^{-2}$& 0.687(39)\\[0.125cm]
${\rm C}_4$&      &  8  & 48   & 2 & 4 &5.2(16)$\cdot 10^{-4}$& 0.944(39)\\[0.125cm]
${\rm C}_5$&      &  12 & 24   & 3 & \{3,6\} &3.3(17)$\cdot 10^{-6}$& 1.052(43)\\[0.125cm]
\hline
\end{tabular}
\caption{Simulation parameters and results. $N_\mathrm{conf}$ is the number of 
configurations of the uppermost level, $N_\mathrm{lev}$ is the number of 
levels and $d$ is the thickness
of the thick time-slice used for the various levels.
The effective mass $M^-$ is given by $-T^{-1} \ln(Z^-/Z)$.
\label{tab:lattices}}
\end{center}
\end{table}

\begin{figure}[ht!]
\begin{center}
\vspace{0.65cm}
\includegraphics[width=10.5cm,height=7cm]{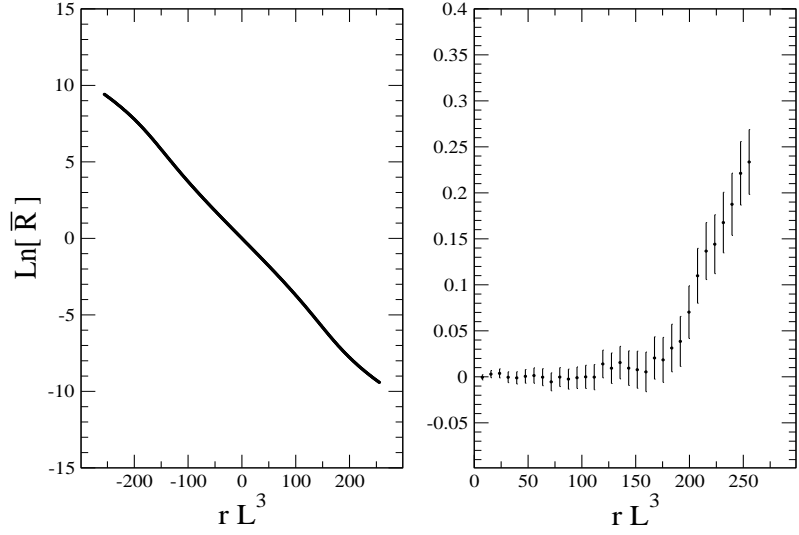}
\caption{Left: the natural logarithm of
$\frac{{\cal Z}_{i-1}}{{\cal Z}_{i}}$ 
is shown as a function of
$r$ (statistical errors are smaller than symbols) for a typical configuration
of the run ${\rm B}_3$.
Right: the sum of the points in the interval $[-r,r]$ is plotted
as a function of $r$ (one each eighth point for visual convenience).}
\label{fig:freeE}
\end{center}
\end{figure}

The natural logarithm of  $\frac{{\cal Z}_{i-1}}{{\cal Z}_{i}}$
for the interpolating systems in Eq.~(\ref{eq:interpol})
is shown as a function of $r$ in the left panel of Fig.~\ref{fig:freeE}
for a typical configuration of the run ${\rm B}_3$. As expected,
its value is of $O(1)$ for each value of $r$. Its almost
perfect asymmetry under $r\rightarrow -r$, however, makes the product of all
the $L^3$ results a quantity of $O(1)$. This impressive cancellation,
which is at work for $T>3$ on all volumes considered,
can be better appreciated in the right panel of the same Figure, where
the sum of the function in the interval $[-r,r]$ is plotted
for a subset of values of $r$. It is the deviation from the exact
asymmetry which flips in sign under a parity transformation of one of
the boundary fields, and forms the signal we are interested in.

\begin{figure}[ht!]
\begin{center}
\includegraphics[width=10.5cm]{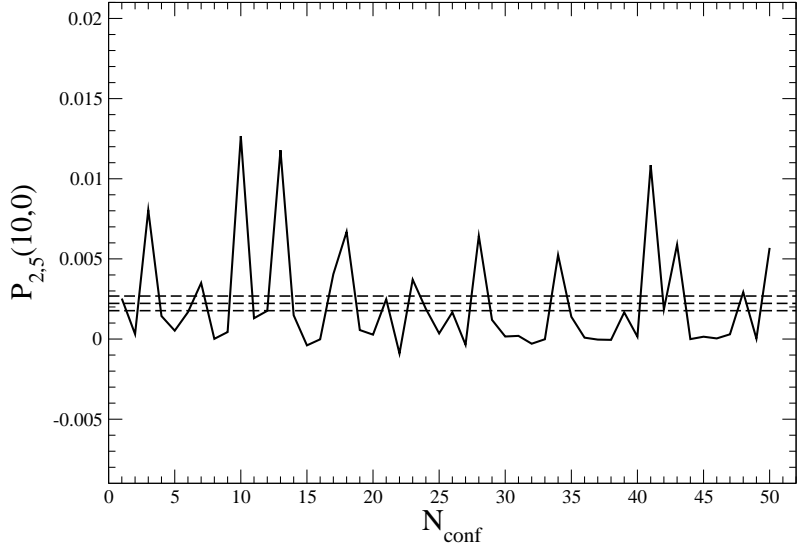}
\caption{Monte Carlo history of the quantity ${{\rm P}}^-_{2,5}[10,0]$ for the run
${\rm A}_5$.}
\label{fig:hist}
\end{center}
\end{figure}

The Monte Carlo history of ${{\rm P}}^-_{2,T/2}[T,0]$
is shown in Figure~\ref{fig:hist} for the lattice ${\rm A}_5$.
The central dashed line corresponds to the average value,
while the other two delimit the one standard deviation region. As expected 
the Monte Carlo history is very regular and each configuration gives an
estimate of the observable which is of the right size. Fluctuations are five times 
the average value at most. We have observed similar Monte Carlo histories also
for the other runs.

\begin{figure}[ht!]
\begin{center}
\includegraphics[,angle=-90,width=10.6cm]{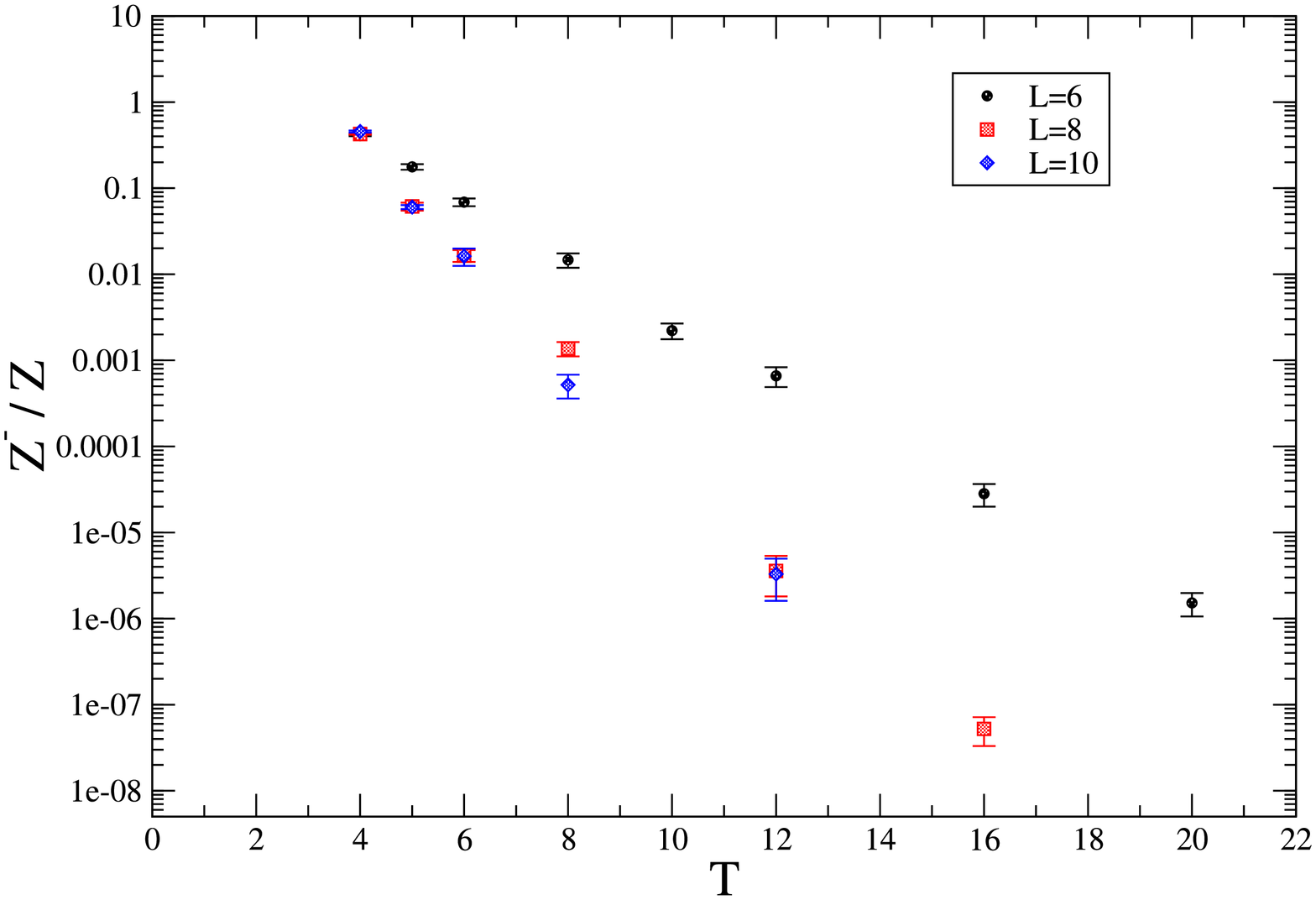}
\caption{The quantity $Z^-/Z$ vs $T/a$.}
\label{fig:corr}
\end{center}
\end{figure}

\begin{figure}[ht!]
\begin{center}
\includegraphics[angle=-90,width=10.4cm]{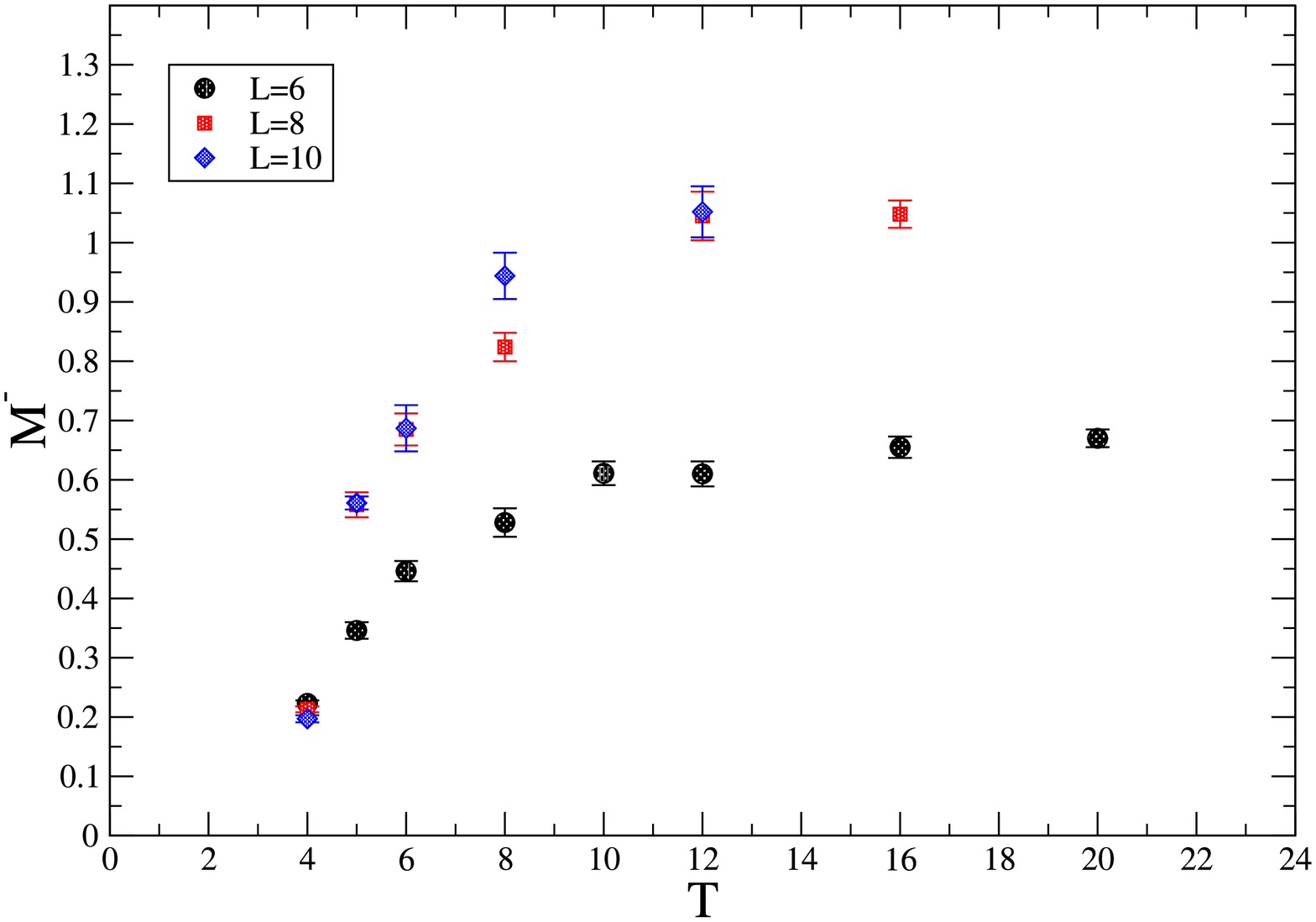}
\caption{The quantity $aM^-$ vs $T/a$.}
\label{fig:meff}
\end{center}
\end{figure}

Finally we show the results for $Z^{-}/Z$ and $aM^{-}$ in Figs.~\ref{fig:corr}
 and~\ref{fig:meff} respectively.
We have been able to follow the exponential decay in the ratio $Z^{-}/Z$ over almost 7 
orders of magnitude. The data at large values of $T/a$ can be used to estimate the 
multiplicity of the first parity odd state, a quantity which is not accessible within
the other approaches. To this end the  precision however has to be
increased, as for now we assume the multiplicity to be one, which justifies the 
definition of the effective mass $M^-$ in Eq.~\ref{eq:M-}.
Figure~\ref{fig:meff} shows that the algorithm works as expected as the error 
on the effective mass 
could be kept constant to the level of a few percent up to a separation of about $3.5$ fm.
It also shows that finite size effects are rather large for lattices of linear size 
around $1$ fm ($L/a=6$) but they become negligible within the present accuracy once 
a size of $1.4$ fm ($L/a=8$) is reached.
We therefore quote $r_0m_{G^-}=3.07(7)$ from $L/a=8$, $T/a=16$ as a preliminary 
result for the mass of the lightest $J^{PC}=0^{-+}$ glueball at a 
lattice resolution of $0.17$ fm with Wilson's gauge action. Given the quite 
large value of the lattice spacing, cutoff effects may
affect this number significantly.
\section{Conclusions and outlook}
\noindent For most of the two-point functions computed on the lattice the noise to signal
ratio grows exponentially with the time separation of the source and the sink.
This disease can be cured 
by imposing the propagation of states with the desired quantum numbers only on each (gauge)
configuration. The algorithm proposed here solves the problem by making use of the 
symmetry properties of the underlying quantum theory.
We have numerically tested the approach in the four-dimensional SU(3) Yang-Mills theory, by
computing the mass of the lightest parity-odd glueball.
For a given precision on the latter the algorithm scales as a power of $T$, the total time 
extent of the lattice, and we have therefore been able to follow an exponential decay
over 7 orders of magnitude and up to separations of $3.5$ fm.
That allows to isolate the contribution of a single state with unprecedented confidence.
We have also studied finite size effect and collected strong indications that, for
the effective mass considered here and within our statistical errors, 
those are negligible for $L>1.4$ fm.
The multiplicity of the state can also be computed using the approach described, and in the
near future we plan to increase the accuracy on its determination, which can be obtained 
only with limited precision by using the data produced so far.
The reduction of systematic uncertainties related to lattice artifacts remains an expensive
task as the algorithm scales roughly as $(L/a)^6$.

The inclusion of other symmetries is straightforward. We have already implemented 
charge-conjugation and tested  it in small volumes, observing basically the same 
efficiency of the integration scheme as for the parity discussed here.
Different symmetry transformations can be actually considered simultaneously and we plan
to include cubic rotations and translations. The mass of the lightest state in any 
sector specified by the quantum numbers $J^{PC}$ could then be computed without
suffering from the exponential problem~\cite{toapp}.

A way to generalize the ideas reported here to systems including fermion 
degrees of freedom is, at present, not known. Among other advantages, such 
an extension would allow to compute the ratio 
between the partition functions at different baryon quantum numbers avoiding
the sign problem, which affects the simulations at finite density.

\end{document}